\documentclass[aps, pre, showpacs, superscriptaddress, reprint, longbibliography]{revtex4-1}
\usepackage{graphicx}
\usepackage{color}
\usepackage{amssymb}
\usepackage{amsmath}
\usepackage{tabularx}
\usepackage{bm}
\usepackage{ltxtable}
\usepackage{natbib} 
\usepackage{longtable}
\newcommand{\be}{\begin{equation}}
\newcommand{\ee}{\end{equation}} 
\newcommand{\bea}{\begin{eqnarray}}
\newcommand{\eea}{\end{eqnarray}}
\newcommand\numberthis{\addtocounter{equation}{1}\tag{\theequation}}

\usepackage{float}

\begin{document}


\title[]{Enstrophy transfers in helical turbulence}

\author{Shubhadeep Sadhukhan}
\email{deep@iitk.ac.in}
\affiliation{Department of Physics, Indian Institute of Technology, Kanpur, Uttar Pradesh 208016, India}

\author{Roshan Samuel}
\email{roshanj@iitk.ac.in}
\affiliation{Department of Mechanical Engineering, Indian Institute of Technology, Kanpur 208016, India}

\author{Franck Plunian}
\email{franck.plunian@univ-grenoble-alpes.fr}
\affiliation{Universit\'e Grenoble Alpes, Universit\'e Savoie Mont Blanc, CNRS, IRD, IFSTTAR, ISTerre, 38000 Grenoble, France}

\author{Rodion Stepanov}
\email{rodion@icmm.ru}
\affiliation{Institute of Continuous Media Mechanics UB RAS, Perm, Russian Federation}
 
\author{Ravi Samtaney}
\email{ravi.samtaney@kaust.edu.sa}
\affiliation{Mechanical Engineering, Division of Physical Science and Engineering, King Abdullah University of Science and Technology - Thuwal 23955-6900, Kingdom of Saudi Arabia}

\author{Mahendra Kumar Verma}
\email{mkv@iitk.ac.in}
\affiliation{Department of Physics, Indian Institute of Technology, Kanpur, Uttar Pradesh 208016}

\date{\today}

\begin{abstract}

In this paper we study the enstrophy transers in helical turbulence using direct numerical simulation.
We observe that the helicity injection does not have significant effects on the inertial-range energy and helicity spectra ($\sim k^{-5/3}$) and fluxes (constants). 
We also calculate the separate contributions to enstrophy transfers via velocity to vorticity  and  vorticity to vorticity channels. There are four different enstrophy fluxes associated with the former channel or vorticity stretching, and one  flux associated with the latter channel or vorticity advection. In the inertial range, the fluxes due to vorticity stretching are larger than that due to advection.  These transfers too are insensitive to helicity injection. 

\end{abstract}

\maketitle

\section{Introduction}
Turbulence is a classic and still open problem in fluid dynamics.
The energetics of homogeneous and isotropic turbulence were explained through the celebrated works of   Kolmogorov~\cite{Kolmogorov:RSA1991a, Kolmogorov:RSA1991b}.
Kolmogorov argued that under statistically steady state, the rate of energy supplied by an external force is equal to the rate at which energy cascades from large to smaller scales which is equal to the rate of energy dissipation. \citet{Obukhov:1949} and \citet{Corrsin:JAP1951} generalized Kolmogorov's theory to  isotropic turbulence  with statistically homogeneous fluctuations of  temperature  field  embedded in a turbulent flow.  The theories of Kolmogorov, Obukhov, and Corrsin, collectively referred to as Kolmogorov-Obukhov theory, predict that the kinetic energy spectrum $E(k)=u_k^2/k  \sim k^{-5/3}$, and the spectrum of temperature fluctuations $E_{\zeta}(k) = \zeta_k^2/k \sim k^{-5/3}$,  where $k = 2\pi/l$ is the wavenumber with $l$ as the length scale, and $u_k$ and $\zeta_k$ are respectively the velocity and temperature fluctuations at wavenumber $k$.  Researchers have also studied the spectra of density and other passive scalars in the same framework. Refer to ~\citet{Lesieur:book:Turbulence} and \citet{Verma:book:BDF} for detailed references.
It is important to note that the effect of helicity is absent in the above-described works. 

Energy and helicity are two inviscid conserved quantities in three-dimensional (3D) turbulent flows.
The effect of helicity in fully developed 3D turbulence has been extensively studied  
\cite{ROBERT:1960,brissaud:POF1973, andre:JFM1977, Polifke:POF1989, waleffe1992, Yokoi:POF1993, Zhou:PRE1993RG, Ditlevsen:POF2001,Chen2003, Lessinnes:PF2011, ChenPRL03, Mininni2011, Pouquet2011,Rodion2018, Biferale2012,Kessar:PRE2015,Sahoo:FDR2018,Avinash:Pramana2006,Teimurazov:CCM2017,Teimurazov:JAM2018}.
\citet{andre:JFM1977} studied the effect of helicity on the evolution of isotropic turbulence at high Reynolds numbers using a variant of Markovian eddy-damped quasi-normal (EDQNM) theory.
\citet{Polifke:POF1989} showed that the energy decay is slowed down by large initial helicity. \citet{waleffe1992} introduced helical decomposition of the velocity field and discussed triadic interactions in helical flows.
\citet{Yokoi:POF1993} reported the effects of helicity in 3D incompressible inhomogeneous turbulence with the help of a two-scale direct-interaction approximation (DIA).
\citet{Zhou:PRE1993RG} showed that helicity does not alter renormalized viscosity.
\citet{Ditlevsen:POF2001,Chen2003}, and \citet{Lessinnes:PF2011} studied helicity fluxes for different types of helical modes.
\citet{ChenPRL03} studied intermittency in helical turbulence.
\citet{Mininni2011} explored the effect of helicity in rotating turbulent flows, while
\citet{Pouquet2011} reported evidences for three-dimensionalization  recovered at small scales.
\citet{Rodion2018} explored the helical bottleneck effect in 3D isotropic and homogeneous turbulence using a shell model.
Maximum helicity states have been studied by \citet{Biferale2012,Kessar:PRE2015}, and \citet{Sahoo:FDR2018}.
Mode-to-mode and shell-to-shell helicity transfers have been studied by \citet{Avinash:Pramana2006}, and \citet{Teimurazov:CCM2017,Teimurazov:JAM2018}.
Helicity may also play a crucial role in magnetohydrodynamics \cite{Verma:PR2004,Alexakis:ApJ2006,Lessinnes:TCFD2009} and dynamo action
\cite{Krause:book:Dynamo,Moffatt:book,Sokoloff:FDR2017,Zeldovich:book:Magneticfield}.

In this paper we derive formulas for  mode-to-mode enstrophy transfers and their associated fluxes.  These quantities arise due to the nonlinear interactions.  In Sec.~\ref{sec-enstrophy} we derive the expressions of mode-to-mode enstrophy transfers and the corresponding fluxes.
In Sec.~\ref{sec-results}, after introducing the helical forcing that we use, the enstrophy fluxes are calculated from direct simulations of 3D turbulence using the pseudo-spectral code TARANG~\cite{Verma:Pramana2013tarang, Chatterjee:JPDC2018}. We conclude in Sec.~\ref{sec-conclusion}.

\section{Enstrophy transfers and fluxes}
\label{sec-enstrophy}
The motion of an incompressible fluid is described by the Navier-Stokes equations
\begin{eqnarray}
    \frac{\partial \mathbf{u}}{\partial t} &=&- (\mathbf{u}\cdot \nabla) \mathbf{u}  -\nabla p +\nu \nabla^2 \mathbf{u}+\mathbf{F}, 
    \label{NSE1}\\
    \nabla \cdot \mathbf{u} &=& 0,
    \label{incompress}
\end{eqnarray}
where $\mathbf{u}$ and $p$ are the velocity and pressure fields, $\nu$ the kinematic viscosity, and $\mathbf{F}$ an external force.

The vorticity $\boldsymbol{\omega}=\nabla \times \mathbf{u}$ obeys the following equation:
\begin{equation}
    \frac{\partial \boldsymbol{\omega}}{\partial t} = (\boldsymbol{\omega}\cdot \nabla) \mathbf{u} -(\mathbf{u}\cdot \nabla) \boldsymbol{\omega} +\nu \nabla^2 \boldsymbol{\omega}+\nabla \times\mathbf{F},
    \label{eq:vorticity}
\end{equation}
where the  first two terms in the right hand side correspond to  vorticity stretching and vorticity advection respectively. In Fourier space the vorticity, which is defined as $\boldsymbol\omega(\mathbf{k}) = i\mathbf{k} \times \boldsymbol u(\mathbf{k})$, satisfies
\begin{eqnarray}
    \frac{d}{dt}\boldsymbol{\omega}(\mathbf{k}) &=&
    i\sum_{\mathbf{p}} (\mathbf{k}\cdot\boldsymbol{\omega}(\mathbf{q}))\mathbf{u}(\mathbf{p})-\left(\mathbf{k}\cdot\mathbf{u}(\mathbf{q})\right)\boldsymbol{\omega}(\mathbf{p})\nonumber\\
     &-& \nu k^2 \boldsymbol{\omega}(\mathbf{k})+ i\mathbf{k} \times \mathbf{F}(\mathbf{k}) ,
    \label{eq:vort_eq}
\end{eqnarray}
where $\mathbf{q}=\mathbf{k}-\mathbf{p}$ and $k=|\mathbf{k}|$.
\subsection{Enstrophy transfers }
Enstrophy, which is defined as $W(\mathbf{k}) = |\boldsymbol{\omega}(\mathbf{k})|^2 /2$,  satisfies
\begin{align*}
    \frac{d}{d t} W(\mathbf{k}) &= \sum_{\mathbf{p}}\biggl[\Im \{(\mathbf{k}\cdot\mathbf{u}(\mathbf{q}))(\boldsymbol{\omega}(\mathbf{p}) \cdot \boldsymbol{\omega}^*(\mathbf{k}))\} \\
                                                                   &-\Im \{(\mathbf{k}\cdot\boldsymbol{\omega}(\mathbf{q}))(\mathbf{u}(\mathbf{p}) \cdot \boldsymbol{\omega}^*(\mathbf{k}))\}\biggr] \\
                                                                   &- 2\nu k^2W(\mathbf{k})+k^2\Re\{\mathbf{u}^*(\mathbf{k}) \cdot \mathbf{F}(\mathbf{k})\} , \numberthis
    \label{eq:enstrophy_flux}
\end{align*}
where ${\bf q=k-p}$. Setting $\textbf{k}'=-\textbf{k}$, the above equation can be rewritten in the form
\begin{align*}
    \frac{d}{dt}W(\mathbf{k'}) &= \sum_{\mathbf{p}} \biggl[S^{\omega\omega}(\mathbf{k'|p,q})+S^{\omega u}(\mathbf{k'|p,q})\biggr] \\
                                                              &- 2\nu k'^2W(\mathbf{k'}) + k'^2\Re\{\mathbf{u}^*(\mathbf{k'}) \cdot \mathbf{F}(\mathbf{k'})\} , \numberthis
                                                              \label{eq:W}
\end{align*}
where  $S^{\omega\omega}(\mathbf{k'|p,q})$ and $S^{\omega u}(\mathbf{k'|p,q})$ are the combined transfers of enstrophy from modes $\mathbf{p}$ and $\mathbf{q}$ to $\mathbf{k'}$, defined as
\begin{eqnarray}
S^{\omega\omega}(\mathbf{k'|p,q})&=&-\Im \{(\mathbf{k'}\cdot\mathbf{u}(\mathbf{q}))(\boldsymbol{\omega}(\mathbf{p}) \cdot \boldsymbol{\omega}(\mathbf{k'}))\}\nonumber \\
                                                          &&- \Im \{(\mathbf{k'}\cdot\mathbf{u}(\mathbf{p}))(\boldsymbol{\omega}(\mathbf{q}) \cdot \boldsymbol{\omega}(\mathbf{k'}))\}, \\
S^{\omega u}(\mathbf{k'|p,q})&=&+\Im \{(\mathbf{k'}\cdot\boldsymbol{\omega}(\mathbf{q}))(\mathbf{u}(\mathbf{p}) \cdot \boldsymbol{\omega}(\mathbf{k'}))\}\nonumber\\
                                                &&+\Im \{(\mathbf{k'}\cdot\boldsymbol{\omega}(\mathbf{p}))(\mathbf{u}(\mathbf{q}) \cdot \boldsymbol{\omega}(\mathbf{k'}))\}
\end{eqnarray}
with ${\bf k'+p+q}=0$. Using the incompressibility condition, we can show that
\bea
S^{\omega\omega}(\mathbf{k'|p,q})+S^{\omega\omega}(\mathbf{p|q,k'})+S^{\omega\omega}(\mathbf{q|p,k'}) &= &0,
\label{eq:enstrophysumzero}
\eea
where $S^{\omega \omega}$ represents the advection of vortices, thus involving enstrophy exchange among the vorticity modes $\mathbf{\boldsymbol{\omega}(k'), \boldsymbol{\omega}(p)}$, and $\mathbf{\boldsymbol{\omega}(q)}$.
On the other hand we have
\bea
S^{\omega u}(\mathbf{k'|p,q})+S^{\omega u}(\mathbf{p|q,k'})+S^{\omega u}(\mathbf{q|k',p}) &\neq& 0,
\label{eq:enstrophysum}
\eea
where $S^{\omega u}$ represents the stretching of vortices, thus either increasing or decreasing the enstrophy $E_{\omega}$.
Incidentally (\ref{eq:enstrophysum}) implies that enstrophy is not a conserved quantity in 3D hydrodynamics.

Now we can split the above transfers into individual contributions from modes $\mathbf{p}$ and $\mathbf{q}$:
\begin{eqnarray}
    S^{\omega\omega}(\mathbf{k'|p,q}) &= &S^{\omega\omega}(\mathbf{k'|p|q}) + S^{\omega\omega}(\mathbf{k'|q|p}) \\
    S^{\omega u}(\mathbf{k'|p,q}) &= &S^{\omega u}(\mathbf{k'|p|q}) + S^{\omega u}(\mathbf{k'|q|p}),
\end{eqnarray}
with
\begin{eqnarray}
    S^{\omega\omega}(\mathbf{k'|p|q}) &=& -\Im [(\mathbf{k'\cdot u(q)}) (\mathbf{\boldsymbol{\omega}(p)\cdot \boldsymbol{\omega}(k')})], \label{eq:mode_to_mode_enstrophyww}\\
    S^{\omega u}(\mathbf{k'|p|q}) &=& +\Im [(\mathbf{k'\cdot \boldsymbol{\omega}(q)}) (\mathbf{u(p)\cdot \boldsymbol{\omega}(k')})]. 
    \label{eq:mode_to_mode_enstrophywu}
\end{eqnarray}
Here $S^{\omega\omega}(\mathbf{k'|p|q})$ and $S^{\omega u}(\mathbf{k'|p|q})$ denote two kinds of mode-to-mode enstrophy transfer, both being from $\mathbf{p}$ to $\mathbf{k'}$ with $\mathbf{q}$ acting as a mediator.
They arise due to  advection and stretching respectively.
Here we note that 
\begin{equation}
S^{\omega\omega}(\mathbf{k'|p|q})=-S^{\omega\omega}(\mathbf{p|k'|q})
\label{eq:Sww}
\end{equation}
due to the incompressibility condition, ${\bf k \cdot u(k)} = 0$; this result is consistent with  (\ref{eq:enstrophysumzero}). Substitution of (\ref{eq:mode_to_mode_enstrophyww}-\ref{eq:mode_to_mode_enstrophywu}) into (\ref{eq:enstrophy_flux}) yields
\begin{align*}
    \frac{d}{dt}W(\mathbf{k'}) &= \sum_{\mathbf{p}} S^{\omega\omega}(\mathbf{k'|p|q}) + \sum_{\mathbf{p}} S^{\omega u}(\mathbf{k'|p|q}) \\
                                                           &- 2\nu k'^2W(\mathbf{k'}) +k'^2\Re\{\mathbf{u}^*(\mathbf{k'}) \cdot \mathbf{F}(\mathbf{k'})\} , \numberthis
    \label{eq:modal_enstrophy}
\end{align*}
and $\mathbf{q=-(k'+p)}$.

\subsection{Enstrophy fluxes}
\begin{figure}
\includegraphics[scale=0.28]{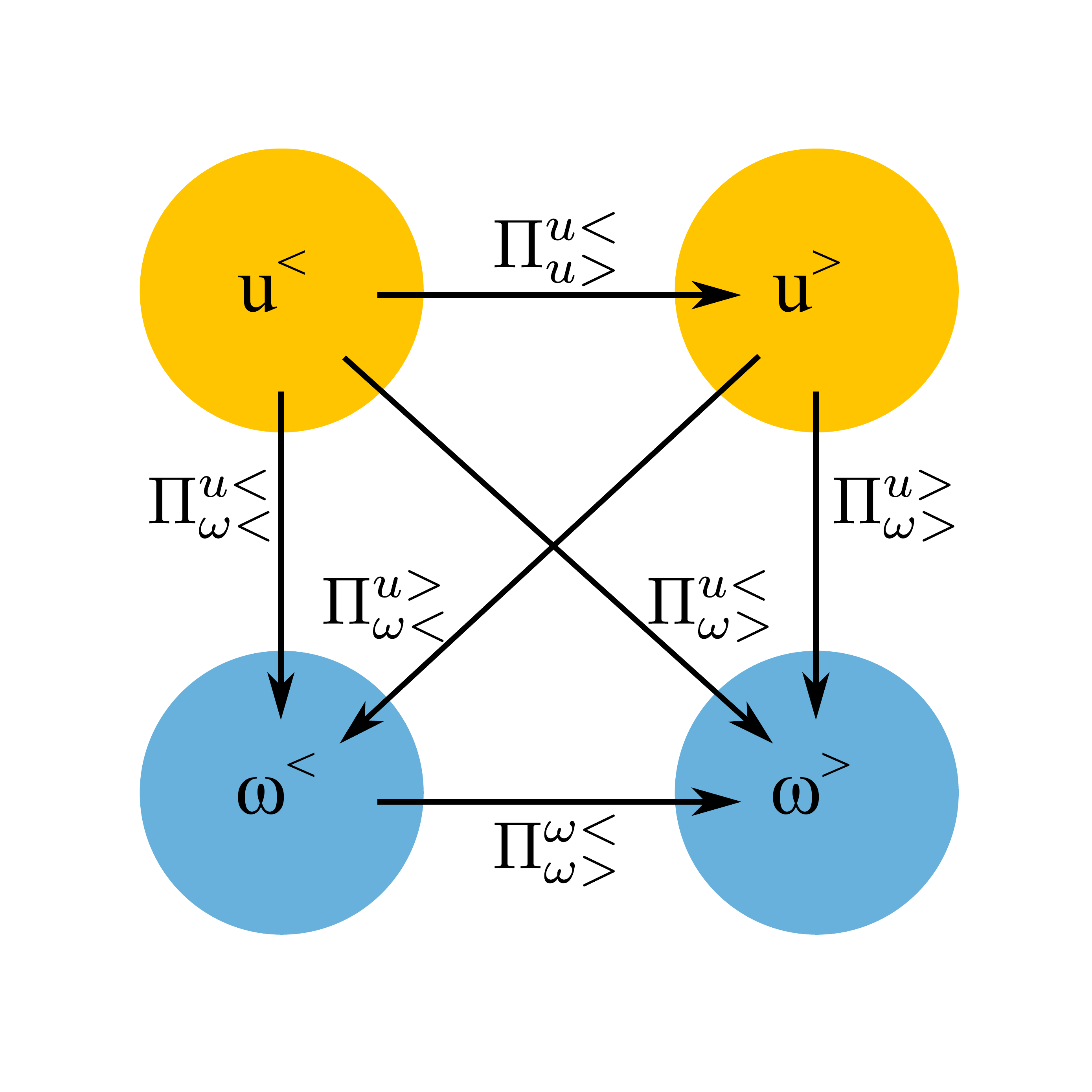}
\caption{(color online) Illustration of the five enstrophy fluxes in addition to the kinetic energy flux $\Pi^{u>}_{u<}\equiv\Pi_E$ which is defined in Eq.~(\ref{eq:Pi_E}).}
\label{fig-schem}
\end{figure}

Summing the previously discussed mode-to-mode enstrophy transfers over $\bf p$ and $\bf k'$ and depending on whether $\bf p$ and $\bf k'$ belong to the sphere of radius $k_0$ or not, we can define the five following enstrophy fluxes  
\begin{eqnarray}
\Pi^{\omega<}_{\omega>}(k_{0})= \sum_{|\mathbf{p}|\le k_{0}} \sum_{|\mathbf{k'}|>k_{0}} S^{\omega \omega}({\bf k'|p|q}) \label{eq:flux1}\\
\Pi^{u<}_{\omega>}(k_{0})= \sum_{|\mathbf{p}|\le k_{0}} \sum_{|\mathbf{k'}|>k_{0}}  S^{\omega u}({\bf k'|p|q})  \label{eq:flux2}\\
\Pi^{u<}_{\omega<}(k_{0})= \sum_{|\mathbf{p}|\le k_{0}} \sum_{|\mathbf{k'}| \le k_{0}}  S^{\omega u}({\bf k'|p|q})\\
\Pi^{u>}_{\omega>}(k_{0})= \sum_{|\mathbf{p}| > k_{0}} \sum_{|\mathbf{k'}|>k_{0}}  S^{\omega u}({\bf k'|p|q})\\
\Pi^{u>}_{\omega<}(k_{0})= \sum_{|\mathbf{p}| > k_{0}} \sum_{|\mathbf{k'}|\le k_{0}}  S^{\omega u}({\bf k'|p|q}),\label{eq:flux5}
\end{eqnarray}
where the superscript and subscript of $\Pi$ represent respectively the giver and the receiver modes,
and the symbols $<$ and $>$ denote respectively the inside and outside of the sphere of radius $k_0$.
Accordingly,
$\Pi^{x<}_{y>}(k_{0})$ denotes the flux of enstrophy 
from all $x$ modes inside the sphere of radius $k_0$ to all $y$ modes outside the sphere.
An illustration of these fluxes is given in Figure \ref{fig-schem}.

Customarily a turbulent flux is associated with some conserved quantity~\cite{Lesieur:book:Turbulence}.  For example, in two-dimensional hydrodynamic turbulence, the energy  and enstrophy fluxes  are connected to the corresponding conservation laws.  The concept of turbulent flux could be generalized to a more complex scenario when a certain quantity is transferred from one field to another. Here, these fluxes are cross transfers among the two fields. Though enstrophy is not conserved, we define enstrophy fluxes of Eqs.~(\ref{eq:flux2}- \ref{eq:flux5}) as enstrophy transfers from large/small scale velocity field to large/small vorticity field.  The enstrophy flux $\Pi^{\omega<}_{\omega>}(k_{0})$ is related to the conservation law related to Eq.~(\ref{eq:enstrophysumzero}). Similar issues arise in magnetohydrodynamic turbulence where fluxes for kinetic and magnetic energies are defined even though kinetic and magnetic energies are not conserved individually.  We define kinetic-to-magnetic energy fluxes as the energy transfers from large/small scale velocity field to from large/small scale magnetic field;  such fluxes are crucial for dynamo action~\citep{Dar:PD2001,Verma:PR2004,Alexakis:ApJ2006}. 

Summing \eqref{eq:modal_enstrophy} over all modes outside the sphere of radius $k_0$, 
we obtain
\begin{eqnarray}
   &&\sum_{|\mathbf{k'}|>k_0}  (\frac{d}{dt}+2\nu k'^2)W(\mathbf{k'})-k'^2\Re\{\mathbf{u}^*(\mathbf{k'}) \cdot \mathbf{F}(\mathbf{k'})\}\nonumber\\ 
   &=&(\Pi^{\omega <}_{\omega >} + \Pi^{u <}_{\omega >} + \Pi^{u >}_{\omega >})(k_0).
    \label{eq:flux_out}
\end{eqnarray}
Performing a sum over all modes inside the sphere of radius $k_0$ yields
\begin{eqnarray}
&&\sum_{|\mathbf{k'}|\le k_0}  (\frac{d}{dt}+2\nu k'^2)W(\mathbf{k'})-k'^2\Re\{\mathbf{u}^*(\mathbf{k'}) \cdot \mathbf{F}(\mathbf{k'})\}\nonumber\\ 
&=&( \Pi^{\omega >}_{\omega <} + \Pi^{u <}_{\omega <} + \Pi^{u >}_{\omega <})(k_0).
\label{eq:flux_in}
\end{eqnarray}

Adding \eqref{eq:flux_out} and \eqref{eq:flux_in} and noticing that (\ref{eq:Sww}) implies $\Pi^{\omega <}_{\omega >}+\Pi^{\omega >}_{\omega <}=0$, we obtain
\begin{eqnarray}
   && \sum_{\mathbf{k'}} (\frac{d}{dt} + 2\nu k'^2) W(\mathbf{k'}) -k'^2\Re\{\mathbf{u}^*(\mathbf{k'}) \cdot \mathbf{F}(\mathbf{k'})\}\nonumber\\ 
    &=&(\Pi^{u <}_{\omega >} + \Pi^{u >}_{\omega >} + \Pi^{u <}_{\omega <} + \Pi^{u >}_{\omega <})(k_0) .
    \label{eq:flux_total}
\end{eqnarray}
As the left hand side of (\ref{eq:flux_total}) does not depend on $k_0$, it implies
\begin{equation}
    \frac{d}{d k_0}\left(\Pi^{u <}_{\omega >} + \Pi^{u >}_{\omega >} + \Pi^{u <}_{\omega <} + \Pi^{u >}_{\omega <}\right) = 0
    \label{eq:flux_deriv}
\end{equation}
which is valid at all times. That is, the following sum
\begin{equation}
\Pi^{u <}_{\omega >}(k_0) + \Pi^{u >}_{\omega >}(k_0) + \Pi^{u <}_{\omega <}(k_0) + \Pi^{u >}_{\omega <}(k_0) = \mathrm{const}
\end{equation}
is independent of $k_0$.

Under a steady state, we obtain
\begin{eqnarray}
   && \sum_{\mathbf{k'}} 2\nu k'^2W(\mathbf{k'}) -\sum_{k_0^F < |\mathbf{k'}| \le k_1^F}k'^2\Re\{\mathbf{u}^*(\mathbf{k'}) \cdot \mathbf{F}(\mathbf{k'})\}\nonumber\\ 
    &=&(\Pi^{u <}_{\omega >} + \Pi^{u >}_{\omega >} + \Pi^{u <}_{\omega <} + \Pi^{u >}_{\omega <})(k_0) ,
    \label{eq:flux_totalsteadystate}
\end{eqnarray}
where the forcing is employed to the wavenumber band $(k_0^F. k_1^F]$. A physical interpretation of the above equation is that the enstrophy injected by the external force is transferred to the enstrophy fluxes and the enstrophy dissipation.

\section{Simulation Method and Results}
\label{sec-results}

The Navier-Stokes equations (\ref{NSE1}) are solved numerically using a fully de-aliased, parallel pseudo-spectral code Tarang~\cite{Verma:Pramana2013tarang, Chatterjee:JPDC2018} with fourth-order Runge-Kutta time stepping.
For de-aliasing purpose, the 3/2-rule has been chosen \cite{Canuto:book:SpectralFluid, Boyd:book:Spectral}.
The viscosity is set to $\nu = 10^{-3}$ and the simulations are performed with a resolution of $512^3$.

\subsection{Helical forcing}
In Fourier space the velocity field satisfies 
\begin{equation}
    \frac{d}{dt}\mathbf{u}(\mathbf{k}) =
   - i\sum_{\mathbf{p}} (\mathbf{k}\cdot\mathbf{u}(\mathbf{q}))\mathbf{u}(\mathbf{p})
    - \nu k^2 \mathbf{u}(\mathbf{k})+ \mathbf{F}(\mathbf{k}) ,
    \label{eq:velocity}
\end{equation}
where $\mathbf{q}=\mathbf{k}-\mathbf{p}$.
Then the energy and helicity, which are defined as
\begin{equation}
E(\mathbf{k})=\frac{|\mathbf{u}(\mathbf{k})|^2}{2},\;\;\; H(\mathbf{k})=\frac{\mathbf{u}(\mathbf{k})\cdot \boldsymbol{\omega}(\mathbf{k})^*}{2},
\end{equation}
satisfy the following equations
\begin{eqnarray}
\frac{d}{dt}E({\bf k'}) &=& -\sum_{\mathbf{p}} \Im \{(\mathbf{k'}\cdot\mathbf{u}(\mathbf{q}))(\mathbf{u}(\mathbf{p}) \cdot \mathbf{u}(\mathbf{k'}))\}\nonumber\\
&-&2 \nu k^2E({\bf k'}) + \Re\{{\bf u^*(k')} \cdot\mathbf{F}(\mathbf{k'})\},\label{eq:Energytime}\\
\frac{d}{dt}H({\bf k'}) &=& \;\;\;\sum_{\mathbf{p}} \Re\{{\mathbf {u(q)}} \cdot ( \boldsymbol{\omega}({\mathbf {p}}) \times  \boldsymbol{\omega}({\mathbf {k'}}) )\}\nonumber\\
&-&2 \nu k^2H({\bf k'}) + \Re\{{\bf \boldsymbol{\omega}^*(k')} \cdot\mathbf{F}(\mathbf{k'})\},\label{eq:Helicitytime}
\end{eqnarray}
where $\mathbf{q}=-(\mathbf{k'}+\mathbf{p})$.

Following \citet{Carati:JoT2006, Carati:PF1995} and \citet{Teimurazov:CCM2017, Teimurazov:JAM2018}, the forcing is taken to be of the form 
\begin{equation}
\mathbf{F(k)}=\alpha \mathbf{u(k)}+\beta \boldsymbol{\omega(k)} 
\label{eq:F}
\end{equation} 
such that 

\begin{equation}
\sum_{k_0^F < |{\bf k}| \le k_1^F} \Re\{{\bf u^*(k)} \cdot\mathbf{F(k)}\}=\epsilon_E,
\label{eq:EF}
\end{equation}
and 
\begin{equation}
\sum_{k_0^F < |{\bf k}| \le k_1^F} \Re\{{\bf {\boldsymbol{\omega}^*}(k)} \cdot\mathbf{F(k)} \}=\epsilon_H.
\label{eq:EH}
\end{equation}
Here, the forcing is applied in the wavenumber band $(k_0^F, k_1^F]$, and $\epsilon_E$ and $\epsilon_{H}$ are the prescribed energy and helicity injection rates. From the above three equations (\ref{eq:F}-\ref{eq:EH}) we deduce that
\begin{equation}
\alpha  =  \frac{1}{2} \frac{W_F \epsilon_E-H_F \epsilon_H}{E_F W_F - H_F^2}, \;\;\;
\beta  =  \frac{1}{2} \frac{E_F \epsilon_H-H_F \epsilon_E}{E_F W_F - H_F^2}, \label{eq:beta}
\end{equation}
where $E_F$,  $H_F$ and $W_F$ are respectively the energy, kinetic helicity and enstrophy  in the forcing band, 
\begin{eqnarray}
(E_F, H_F,W_F)&=&\sum_{k_0^F < |{\bf k}| \le k_1^F} (E({\bf k}),H({\bf k}),W({\bf k})). 
\end{eqnarray}
In our simulations, the forcing band corresponds to $k_0^F=2$ and $k_1^F=3$. 
The main advantage of using the forcing given by (\ref{eq:F}-\ref{eq:beta}) is that the injection rates of energy and helicity can be set independently.
The realizability condition $|H({\bf k})| \le k E({\bf k})$ directly comes from the definitions of energy and helicity and is therefore always satisfied, whatever the values of $\epsilon_E$ and $\epsilon_H$. For example even in the case $\epsilon_E \ll \epsilon_H$, which was studied by \citet{Kessar:PRE2015}, the realizability condition is satisfied, reaching a state close to the maximal helical state $|H({\bf k})| = k E({\bf k})$. 

\begin{figure}
\includegraphics[scale=0.45]{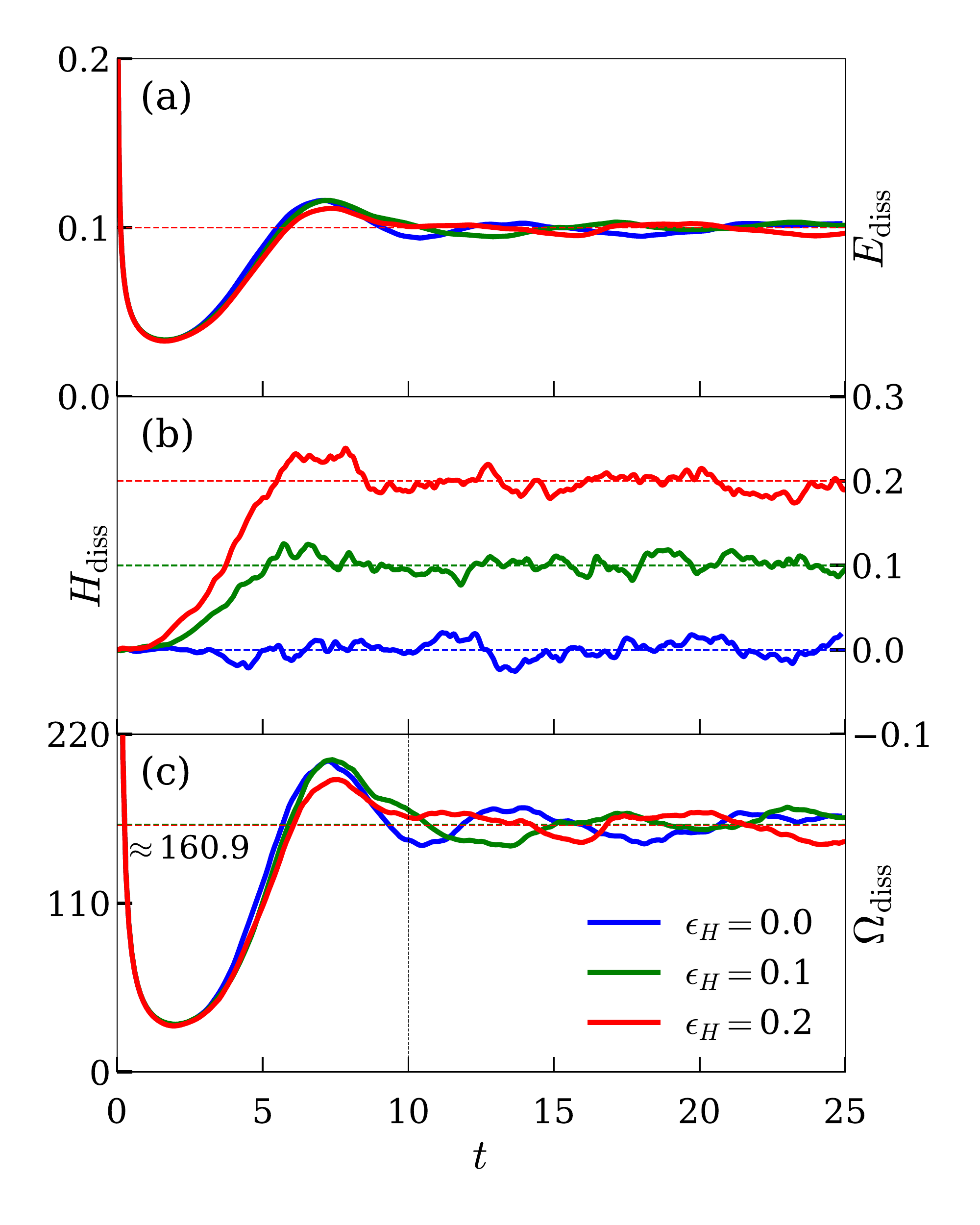}
\caption{(color online) Time evolution of (a) energy dissipation, (b) helicity dissipation, and (c) enstrophy dissipation for the same energy injection rate $\epsilon_E=0.1$ and three different helicity injections $\epsilon_H=0, 0.1, 0.2$ .
For $t\ge10$ a steady state is reached with a mean enstrophy dissipation about $160.9$.}
\label{fig_evolution}
\end{figure}

In Fig.~\ref{fig_evolution} the time evolution of (a) the total energy dissipation $2\nu \sum\limits_{\mathbf{k}} k^2 E(\mathbf{k})$, (b) the total helicity dissipation $2\nu \sum\limits_{\mathbf{k}} k^2 H(\mathbf{k})$, and (c) the total enstrophy dissipation $2\nu \sum\limits_{\mathbf{k}} k^2 W(\mathbf{k})$ are plotted for the same energy injection rate $\epsilon_E=0.1$ and three helicity injections $\epsilon_H=0, 0.1, 0.2$.   We observe that a statistical steady state is reached for $t\ge 10$.
In this steady state we find that, on an average,  $2\nu \sum\limits_{\mathbf{k}} k^2 E(\mathbf{k})\approx \epsilon_E$ and $2\nu \sum\limits_{\mathbf{k}} k^2 H(\mathbf{k})\approx \epsilon_{H}$ which is consistent with the prescribed forcing.

\subsection{Energy and helicity spectra and fluxes}
\begin{figure}
\includegraphics[scale=0.265]{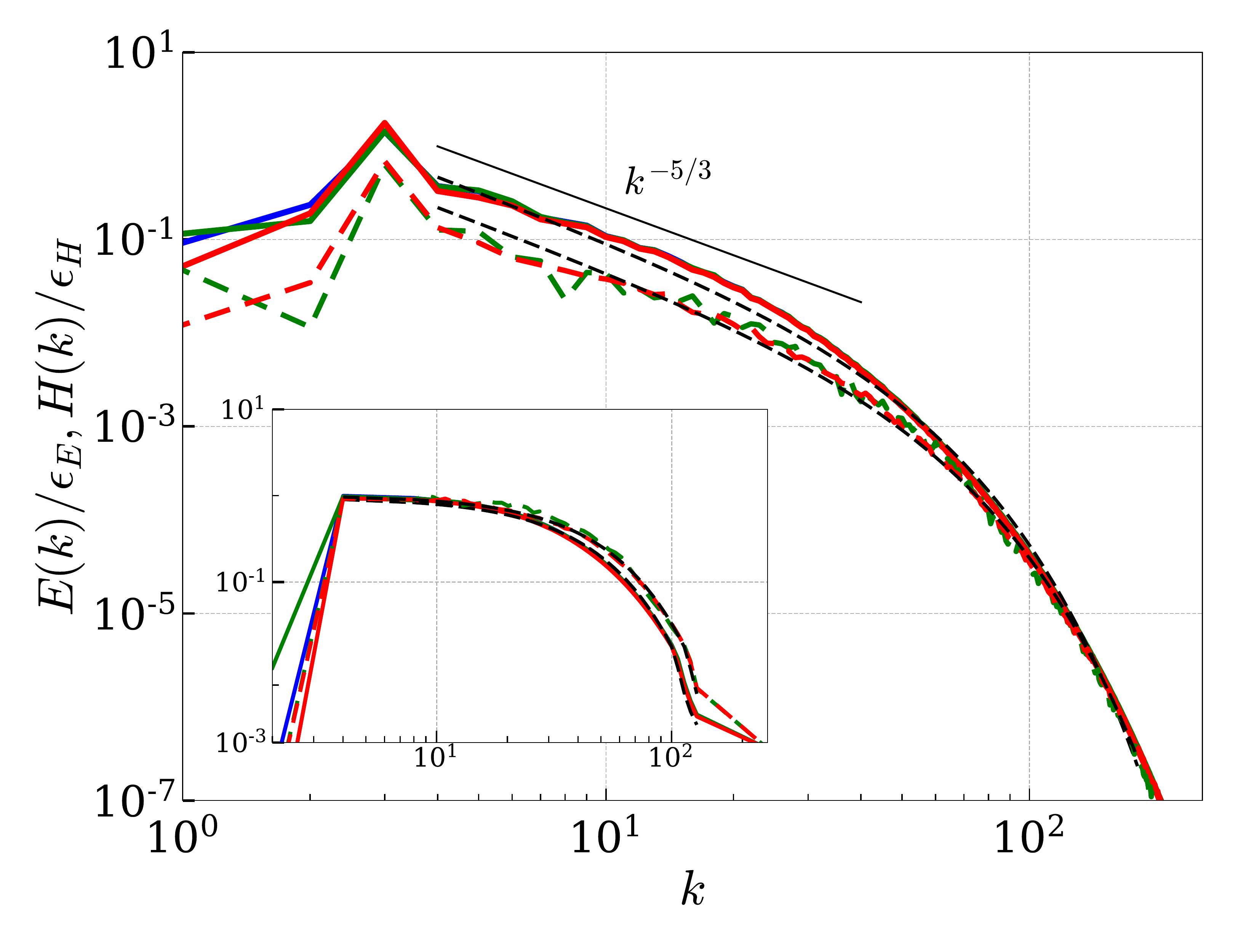}
\caption{(color online) Normalized energy spectra (solid curves) and helicity spectra (dashed curves) with the same color code as in Fig.~\ref{fig_evolution}.
         In the inset, normalized energy fluxes (solid curves) and helicity fluxes  (dashed curves).
         The black dashed curves correspond to the analytical formula (\ref{eq:Pao1}-\ref{eq:Pao4}) .}
\label{energy_spectra}
\end{figure}

From (\ref{eq:Energytime}) and (\ref{eq:Helicitytime}) and following steps analogous to (\ref{eq:enstrophy_flux}-\ref{eq:mode_to_mode_enstrophywu}), the following expression for mode-to-mode energy and helicity transfers can be derived:
\begin{eqnarray}
S^{E}(\mathbf{k'|p|q}) &=& -\Im \{(\mathbf{k'}\cdot\mathbf{u}(\mathbf{q}))(\mathbf{u}(\mathbf{p}) \cdot \mathbf{u}(\mathbf{k'}))\}, \\
S^{H}(\mathbf{k'|p|q}) &=& +\Re\{{\mathbf {u(q)}} \cdot ( \boldsymbol{\omega}({\mathbf {p}}) \times  \boldsymbol{\omega}({\mathbf {k'}}) )\},
\end{eqnarray}
leading to the following flux definitions
\begin{equation}
\Pi_{E,H}(k_0) = \sum_{|{\bf k'}|>k_0}   \sum_{|{\bf p}|<k_0} S^{E,H}(\mathbf{k'|p|q}) .
\label{eq:Pi_E}
\end{equation}
In Fig.~\ref{energy_spectra}, the energy and helicity spectra, normalized by respectively $\epsilon_E$ and $\epsilon_H$, show a spectral scaling close to  $k^{-\frac{5}{3}}$
in the inertial range, for the three injection rates of helicity $\epsilon_H=0, 0.1, 0.2$. The normalized fluxes shown in the inset are approximately flat in the inertial range, and again insensitive to $\epsilon_H$.  

In Fig.~\ref{energy_spectra} the black dashed curves correspond to the
analytical formula for spectra and fluxes derived by \citet{Pao:PF1965} for the energy and that are here extended to helicity.
They take the following form
\begin{eqnarray}
\frac{E(k)}{\epsilon_E}&=& \frac{K_E}{\epsilon_E^{1/3}}k^{-5/3}\textrm{exp}\left(-\frac{3}{2} K_E(k/k_d)^{4/3}\right),\label{Ek_nh} \label{eq:Pao1}\\
\frac{H(k)}{\epsilon_H}&=& \frac{K_H}{\epsilon_E^{1/3}}k^{-5/3}\textrm{exp}\left(-\frac{3}{2} K_H(k/k_d)^{4/3}\right), \label{Ek_h}\\
 \frac{\Pi_E(k)}{\epsilon_E} &=&  \textrm{exp}\left(-\frac{3}{2} K_E(k/k_d)^{4/3}\right),\label{Pik_nh}\\
\frac{\Pi_H(k)}{\epsilon_H}&=& \textrm{exp}\left(-\frac{3}{2} K_H(k/k_d)^{4/3}\right),\label{eq:Pao4}
\label{Pi_h}
\end{eqnarray}
where $K_E$ is Kolmogorov's constant, and $K_H$  is another non-dimensional constants, and $k_{d}=(\epsilon_E/\nu^{3})^{1/4}$ is the Kolmogorov's wavenumber. Note however that Pao's spectrum slightly  overestimates the dissipation range spectrum~\citep{Verma:FD2018}, and we expect similar discrepancies for the kinetic helicity that may show up in high-resolution simulations.  These issues may be related to intermittency and enhanced dissipation due to bottleneck effect~\citep{Falkovich:PF1994, Verma:JPA2007}.


\subsection{Results of enstrophy fluxes}
\label{sec-fluxes}
\begin{figure}
\includegraphics[scale=0.27]{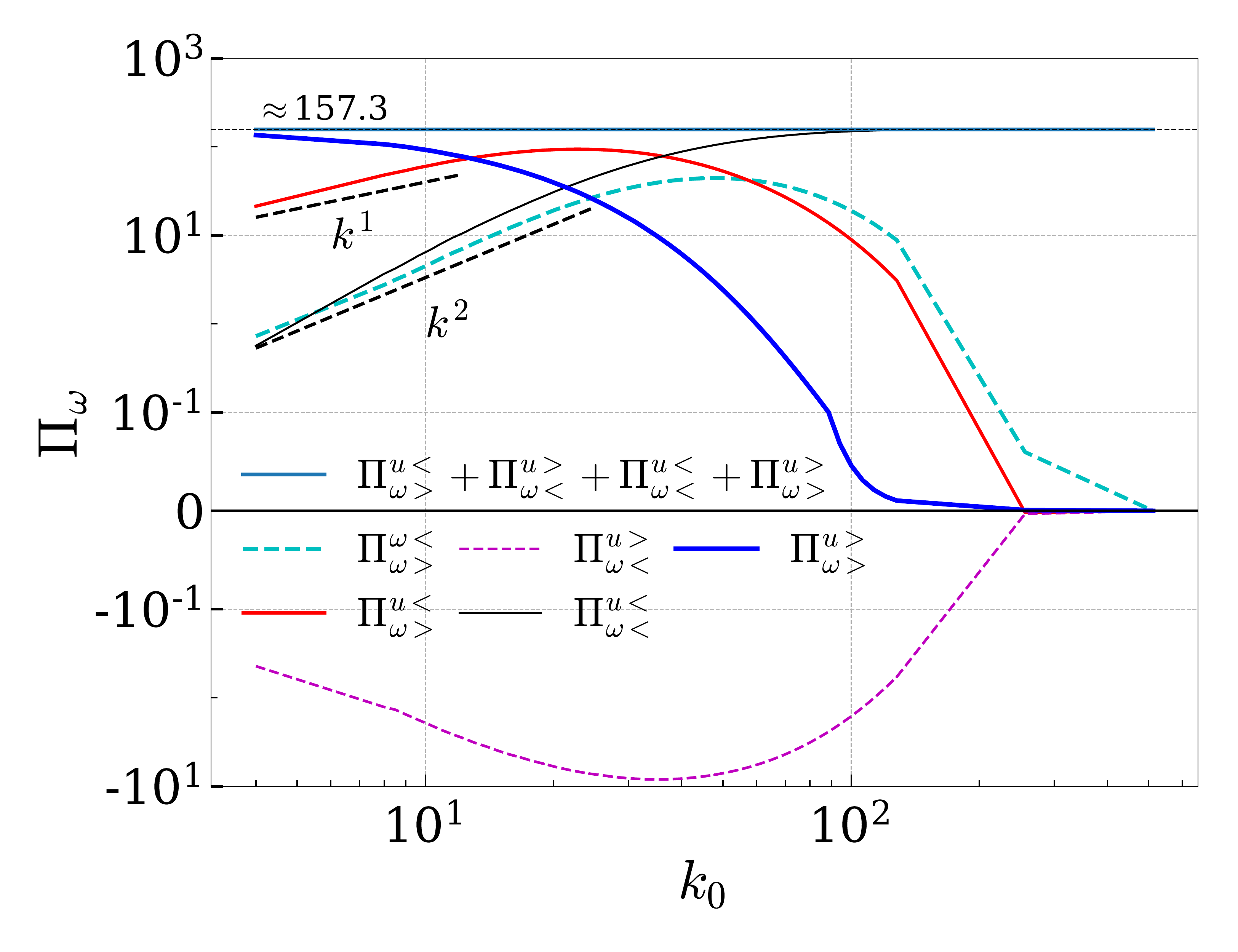}
\caption{(color online) The five enstrophy fluxes given in (\ref{eq:flux1}-\ref{eq:flux5}), and the sum $\Pi^{u <}_{\omega >}+\Pi^{u >}_{\omega <}+\Pi^{u <} _{\omega <}+\Pi^{u >} _{\omega >}$ are plotted for $\epsilon_E=0.1$ and $\epsilon_{H}=0$.}
\label{fig-piw}
\end{figure}

The enstrophy fluxes given in (\ref{eq:flux1}-\ref{eq:flux5}) have been calculated for $\epsilon_{H}=0, 0.1, 0.2$.  They are found to be insensitive to $\epsilon_{H}$. Therefore only the results corresponding to $\epsilon_{H}=0$ are presented in Fig.~\ref{fig-piw}.  
The manifestation of this independence is consistent with many numerical results which showed that the forward flux of kinetic energy is also insensitive to kinetic helicity injection at large scale \cite{2018PhR...767....1A}. Crucial changes can be expected when significant imbalance between helical modes of different signs is imposed by helicity forced over a wide range of scales \cite{Kessar:PRE2015}.  

The flux $\Pi^{\omega <}_{\omega >}$ is positive suggesting a direct cascade of enstrophy. In the inertial range it obeys a
$k^2$ scaling law. However we observe that in the inertial range, $\Pi^{u <}_{\omega >}$ is larger than $\Pi^{\omega <}_{\omega >}$, showing that the enstrophy flux due to stretching is larger than the one due to advection.  We also observe that $\Pi^{u >}_{\omega <}$ is always negative, indicating that the small-scale  velocity fluctuations squeeze  (not stretch) the large scale vorticity.  It also obeys a $k^2$ scaling law in the inertial range.  

The flux $\Pi^{u <}_{\omega <}(k_0)$ is an accumulated sum of all the enstrophy transferred from $u^<$ to $\omega^<$ up to the radius $k_0$. All the wavenumber shells have positive sign for these transfer, hence $\Pi^{u <}_{\omega <}(k_0)$ is an increasing function of $k_0$.  Note that $\Pi^{u <}_{\omega <}(\infty)$ is the net enstrophy transfer.   The complementary flux, $\Pi^{u >}_{\omega >}(k_0)$, has an opposite behaviour; that is, it decreases with $k_0$.  All the fluxes other than $\Pi^{u <}_{\omega <}(k_0)$ vanish for $k_0 > k_d$ because the fluctuations vanish in this range.

It is found that $(\Pi^{u <}_{\omega >}+\Pi^{u >}_{\omega <}+\Pi^{u <} _{\omega <}+\Pi^{u >} _{\omega >})(k_0)\approx 157.3$, independently of $k_0$ to a precision of about $10^{-10}$. This value is comparable to the enstrophy dissipation which is about 160.9 (Fig.~\ref{fig_evolution}); the difference between the two quantities is  $\sum\limits_{{k_0^F < |{\bf k'}| \le k_1^F}}|k'|^2\Re\{\mathbf{u}^*(\mathbf{k'}) \cdot \mathbf{F}(\mathbf{k'})\} \approx 3.6.$; these computations are consistent with   (\ref{eq:flux_totalsteadystate}). 

\section{Conclusions}
\label{sec-conclusion}
Using direct numerical simulations and varying the injection rate of helicity we find that injecting helicity does not change the results in terms of energy and helicity spectra and fluxes, and also in terms of enstrophy fluxes. The energy and helicity spectra and fluxes follow rather well the formula of  \citet{Pao:PF1965} that we extended to helicity. Of course, some correction due to intermittency and bottleneck effect should be added, but these issues are beyond the scope of the present paper.  These results are consistent with the predictions that the inertial range properties of hydrodynamic turbulence is independent of kinetic helicity~\cite{Zhou:PRE1993RG, Avinash:Pramana2006}.

The main objective of this paper is to introduce the derivation of mode-to-mode enstrophy transfers using which we compute the enstrophy fluxes.  There are five different enstrophy fluxes. Four of them, $\Pi^{u<}_{\omega <}$, $\Pi^{u<}_{\omega >}$, $\Pi^{u>}_{\omega <}$, $\Pi^{u>}_{\omega >}$, correspond to vorticity stretching ,  
the fifth one $\Pi^{\omega<}_{\omega >}$ is due to the advection of vorticity by the flow.
It is remarkable that in the inertial range $\Pi^{u<}_{\omega >}$ is larger than $\Pi^{\omega<}_{\omega >}$
implying that the enstrophy flux due to vorticity stretching is larger than the one due to advection of vorticity.
The sum of the first four fluxes corresponds to the contribution from all velocity modes to all vorticity modes and therefore should not depend on the wavenumber, that we confirm numerically with a precision better than $10^{-10}$.\\

\begin{acknowledgments}
Our numerical simulations have been performed on Shaheen II at {\sc Kaust} supercomputing laboratory, Saudi Arabia, under the project k1052.
This work was supported by the research grants PLANEX/PHY/2015239 from Indian Space Research Organisation, India, and by the Department of Science and Technology, India (INT/RUS/RSF/P-03).
\end{acknowledgments}

\end{document}